# Bidirectional teleportation for underwater quantum communications

Sundaraja Sitharama Iyengar,* K. J. Latesh Kumar, and Mario Mastriani

*School of Computing & Information Sciences, Florida International University, 11200 S.W. 8th Street, Miami, FL 33199, USA*

## Abstract

In this work, we evaluate the performance of a bidirectional teleportation protocol on an IBM-Q's quantum processor of six or more qubits. If the experiment is successful, we will implement this protocol between two submerged nuclear submarines on opposite sides of the ocean thanks to a satellite that generates and distributes entangled pairs, as well as transmits optical bits of disambiguation between both submarines.

ORCID Id:

S.S. Iyengar: 0000-0003-3203-833X

K.J. Latesh Kumar: 0000-0003-0027-9972

M. Mastriani: 0000-0002-5627-3935

---

*Electronic address: iyengar@cis.fiu.edu

## I. INTRODUCTION

Since the first implementation of the quantum teleportation protocol of Bennett et al [1], by Bouwmeester et al [2], which is basically a one-way version (i.e., forward) in terms of Information direction, an efficient bidirectional version that can be used in a practical context has been feverishly sought. On the other hand, all the implementations of the quantum teleportation protocol [2-5] have been carried out on optical circuits [6], because this protocol implies the use of the *if-then-else* statement, which cannot be implemented on a quantum processing unit (QPU) based on the superconductors technology, such as those belonging to the IBM Q [7] family of processors.

If we also think that a practical application for a bidirectional teleportation protocol could be the simultaneous exchange of keys between two nuclear submarines submerged in the ocean thanks to a quantum satellite, the only possible implementation for said protocol is thanks to entangled photons, since the quantum technology based on superconductors completely loses the quantum attribute at environment temperature by the direct action of decoherence in a non-adiabatic context. Therefore, and by virtue of the two problems raised, i.e., the impossibility of implementing the *if-then-else* statement, and its exclusive confinement in a laboratory with the necessary low temperatures, we can conclude that QPUs based on superconductor technology is completely excluded from the proposed application, i.e., underwater quantum communications, being exclusively reserved the implementation of such protocol on an optical circuit [6] based on entangled photons.

However, if what we are interested is in the study of the viability of the protocol as well as its satellite configuration from the operational point of view and implementations on a QPU like those of IBM Q [7] would be very well received, considering that the cost of an optical laboratory for this type of experiment would be hundreds of thousands of dollars (even much more), while the implementations on IBM Q [7] QPUs are freely accessible via the web. Therefore, the only thing left for us to solve is to find a bidirectional protocol that does not require the *if-then-else* statement. This protocol exists and will be used in the following sections on different IBM Q [7] environments.

Summing-up, we confirm that the focus of this work has exclusively to do with the operational viability of the protocols and their satellite configuration, leaving aside all technical aspects related to the practical implementation of the underwater quantum communications solution. In particular, in later sections, we will mention that although the submarines are submerged, they send a buoy to the surface that will fulfill two functions: receiving the entangled photons sent by the quantum satellite, as well as sending and receiving signals (electromagnetics or optical) to and from such satellite, respectively, where such signals represent the bits of disambiguation required by the protocol [1-5].

In consequence, two technical problems automatically appear:

a) the buoy wobbling on the surface, even with a calm sea, preventing any tracking system via a laser beam, which will automatically force the use of a spotlight of entangled photons with a very closed solid angle (i.e., well confined on the buoy) so that a third one does not receive these entangled photons and end up being part of the quantum channel, and

b) the extense footprint of the electromagnetic classic channel on the sea surface emitted from the satellite antenna is used to transmit and receive the disambiguation bits. This footprint can be intercepted by a third silent, submerged, and pre-existing submarine in the vicinity of the area of some of the allied submarines, decoding and altering the mentioned bits. In these circumstances, the security of the message would not be at stake, but the integrity of it, i.e., the third one will not be able to decode the message but neither will the allied receiver. Therefore, as we will see in Section III, we will use a Skyloom's satellite type Uhura [8] which will allow the disambiguation bits to be sent by means of an optical link with a solid angle as closed as that used for entangled photons, which will disable the interception intention of any third party in the area.

Finally, if the entire link depends on buoys (however small and compact they are) on the surface, the quantum communication system will not be completely underwater, although it will come quite close.

The rest of this paper is organized as follows. The relevant background concepts relating to underwater quantum communications, which implies: analysis of the qubits to be employed in this particular experiment, bidirectional teleportation, and satellite quantum link; are summarized in Section II for facilitating the presentation of the implementation on a quantum processing unit (QPU) of IBM Q [7] in Section III. Section IV contents our conclusions.

## II. UNDERWATER QUANTUM COMMUNICATIONS

### A. Setup

Figure 1 shows the Bloch's sphere, where we can see its two poles:

$$North\ pole = Spin\ up = |0\rangle = \begin{bmatrix} 1 \\ 0 \end{bmatrix}, \text{ and} \tag{1}$$

$$South\ pole = Spin\ down = |1\rangle = \begin{bmatrix} 0 \\ 1 \end{bmatrix}. \tag{2}$$

These poles are called Computational Basis States (CBS). All pure state can be represented on the Bloch's sphere [9-11] of Fig. 1 as a superposition of both CBS $\{|0\rangle, |1\rangle\}$, resulting in a wave-function like the following,

$$|\psi\rangle = \alpha|0\rangle + \beta|1\rangle = \alpha\begin{bmatrix} 1 \\ 0 \end{bmatrix} + \beta\begin{bmatrix} 0 \\ 1 \end{bmatrix} = \begin{bmatrix} \alpha \\ \beta \end{bmatrix} \tag{3}$$

where $|\alpha|^2 + |\beta|^2 = 1$, such that $\alpha \wedge \beta \in \mathbb{C}$ of a Hilbert's space [9]. Specifically, the most practical version of the wave-function in terms of the angles of Fig. 1, will be,

$$|\psi\rangle = cos\frac{\theta}{2}|0\rangle + e^{i\phi} sin\frac{\theta}{2}|1\rangle \tag{4}$$

where $0 \leq \theta \leq \pi$, and $0 \leq \phi < 2\pi$ [9]. Besides, Equations (3) and (4) are identical for $\alpha = cos(\theta/2)$ and $\beta = e^{i\phi} sin(\theta/2)$. The angles $\theta$ and $\phi$, along with the radius of the sphere $r$, define a point on the unit three-dimensional sphere, as shown in Fig. 1, with a Cartesian,

$$|\psi\rangle \equiv \begin{bmatrix} z = cos(\theta/2) = \alpha \\ x = cos\phi\ sin(\theta/2) = \beta_x \\ y = i\ sin\phi\ sin(\theta/2) = \beta_y \end{bmatrix} \tag{5}$$

or polar,

$$|\psi\rangle \equiv \begin{bmatrix} r \\ \theta \\ \phi \end{bmatrix} \tag{6}$$

representation, being $r = 1$, where $\beta_x$ is the projection of the wave-function onto $x$ axis, while $\beta_y$ is the projection of the wave-function onto $y$ axis. Then, given a generic qubit on Hilbert space $H_{2\times 1}$ as that of Eq.(3), its density matrix in $H_{2\times 2}$ will be,

$$\rho_{|\psi\rangle} = |\psi\rangle\langle\psi^*| = \begin{bmatrix} \alpha \\ \beta \end{bmatrix}\begin{bmatrix} \alpha^* & \beta^* \end{bmatrix} = \begin{bmatrix} |\alpha|^2 & \alpha\beta^* \\ \beta\alpha^* & |\beta|^2 \end{bmatrix}, \qquad (7)$$

where (•)* means complex conjugate of (•). On the other hand, the elements on the main diagonal of the density matrix of Eq.(7) will represent the probabilities or outcomes obtained as a consequence of the quantum measurement process [12-15] of qubits:

$$\text{Probability of } |0\rangle \, (\text{Po}|0\rangle) = \rho_{|\psi\rangle_{(1,1)}} = |\alpha|^2, \text{ and} \qquad (8a)$$

$$\text{Probability of } |1\rangle \, (\text{Po}|1\rangle) = \rho_{|\psi\rangle_{(2,2)}} = |\beta|^2. \qquad (8b)$$

From now on, we will use, in all experiments, the next two interesting qubits:

$$THTH|0\rangle = \begin{bmatrix} 1 & 0 \\ 0 & exp(\frac{i\pi}{4}) \end{bmatrix} \tfrac{1}{\sqrt{2}}\begin{bmatrix} 1 & 1 \\ 1 & -1 \end{bmatrix}\begin{bmatrix} 1 & 0 \\ 0 & exp(\frac{i\pi}{4}) \end{bmatrix}\tfrac{1}{\sqrt{2}}\begin{bmatrix} 1 & 1 \\ 1 & -1 \end{bmatrix}\begin{bmatrix} 1 \\ 0 \end{bmatrix}, \text{ and} \qquad (9a)$$

$$XTHTH|0\rangle = \begin{bmatrix} 0 & 1 \\ 1 & 0 \end{bmatrix}\begin{bmatrix} 1 & 0 \\ 0 & exp(\frac{i\pi}{4}) \end{bmatrix}\tfrac{1}{\sqrt{2}}\begin{bmatrix} 1 & 1 \\ 1 & -1 \end{bmatrix}\begin{bmatrix} 1 & 0 \\ 0 & exp(\frac{i\pi}{4}) \end{bmatrix}\tfrac{1}{\sqrt{2}}\begin{bmatrix} 1 & 1 \\ 1 & -1 \end{bmatrix}\begin{bmatrix} 1 \\ 0 \end{bmatrix}, \qquad (9b)$$

whose main characteristics can be seen in Table I. They were chosen due to their equidistant positions (each in the antipodes of the other) on the Bloch sphere, for being an excellent example of overlapping of both CBSs, and about which there are precedents in the literature [16] of their implementation on the two platforms used in this work, i.e., Quirk [17], and IBM Q [7].

### B. Bidirectional teleportation

In this subsection, the most practical version of the bidirectional quantum teleportation protocols [18-21] is presented, which is also known as teleportation-swapping protocol. This one allows interchange keys, messages, or states, between Alice and Bob in a simultaneous way. Essentially, it makes a swapping or *quid pro quo* procedure between two qubits, as we can see in Fig. 2(c) on the Quirk [17] platform. However, this protocol uses the *if-then-else* statement, which cannot be implemented on a quantum processing unit (QPU) like those of the IBM Q [7] family. Therefore, we will resort to the equivalent protocol of Fig. 2 (d), which will give us the same outcomes than the original protocol but it does not use the *if-then-else* statement. The protocol of Fig. 2(d) is not apply in practice due to the classic channel would transmit a version of the qubits to be teleported, thus exposing all the security if said channel were intercepted by a third party, while the protocol of Fig. 2(c) only transmits disambiguation bits on the classic channel from which it is impossible to reconstruct the qubit to be teleported, however, as already we have mentioned, it cannot be implemented on a QPU. Therefore, in Sec. III, we will use the protocol of Fig. 2(d) with the sole objective of studying the operational feasibility of the satellite configuration that will be discussed in that section. In fact, this is all we can do with a QPU technology that is based on superconductors [7], given that if we tried to get superconductors out of their adiabatic and refrigerated environment, they would fall prey to decoherence [13] with the consequent death of entanglement [22-24] and any possibility of teleportation [1-5].

Finally, we will describe, in order, the complete content of Fig. 2, where all its elements are implemented on the Quirk simulator [17], which incorporates the following metrics: Bloch's sphere (BS), density matrix (DM), and Chance of being ON if measured (Po|1>), among others. These metrics must be interpreted as follows:

- BS: Bloch's sphere with an orientation according to the projections of a qubit on axis *x*, *y*, and *z*.
- DM: Density matrix, where the radius of the green circle in every quadrant represents the modulus of the correspondent element, the orientation of the black segment means the angle (or phase) of the correspondent element, and the level of green in the elements whose positions are 00 and 11 means: Probability of |0> (Po|0>), and Probability of |1> (Po|1>), respectively.
- Po|1>: Probability of |1> has its own icon in Quirk [17], which allows us to automatically deduce the Probability of |0> (Po|0>) by means of a simple subtraction: Po|0> = 1 - Po|1>, which is absolutely concomitant with the values of Table I.

Then, for Fig. 2 is formed by the following elements:
a) the qubit THTH|0>, which will have the following characteristics on Quirk [17]:
- *Bloch sphere (BS) representation of local state:*

    $r = +1.0000$, $\varphi = -45.00°$, $\theta = +45.00°$

    $x = +0.5000$, $y = -0,5000$, $z = +0.7071$

- *Density matrix (DM):*

  00: Probability of |0> (decimal 0) = 85.3553 %

  01: Coupling of |0> to <1| (decimal 0 to 1) = +0.250000+0.250000i

  10: Coupling of |1> to <0| (decimal 1 to 0) = +0.250000-0.250000i

  11: Probability of |1> (decimal 1) = 14.6447 %

- *Chance of being ON if measured (Po|1>):*

  Po|1> = 14.64466 %

b) the qubit XTHTH|0>, which will have the following characteristics on Quirk [17]:

- *Bloch sphere (BS) representation of local state:*

  $r = +1.0000, \varphi = +45.00°, \theta = +135.00°$

  $x = +0.5000, y = +0,5000, z = -0.7071$

- *Density matrix (DM):*

  00: Probability of |0> (decimal 0) = 14.6447 %

  01: Coupling of |0> to <1| (decimal 0 to 1) = +0.250000-0.250000i

  10: Coupling of |1> to <0| (decimal 1 to 0) = +0.250000+0.250000i

  11: Probability of |1> (decimal 1) = 85.3553 %

- *Chance of being ON if measured (Po|1>):*

  Po|1> = 85.3553 %

c) the bidirectional teleportation protocol, with *if-then-else* statement, on Quirk [17] between Alice (red on sea) and Bob (blue on sea) thanks to a quantum satellite (green), where the qubit THTH|0> to be teleported begins in q[0] on the left of the red sector, and finish on q[5] on the right of the blue sector, indicating that the teleportation from Alice to Bob was successful, and

d) an alternative bidirectional teleportation protocol, without *if-then-else* statement, on Quirk [17] between Alice (red on sea) and Bob (blue on sea) thanks to a quantum satellite (green), where the qubit XTHTH|0> to be teleported begins in q[5] on the left of the blue sector, and finish on q[0] on the right of the red sector, indicating that the teleportation from Bob to Alice was successful too.

### C. Satellite quantum link

Figure 3(a) shows a configuration constituted by two allied submarines (A and B in gray), a third, silent, submerged and pre-existing not ally submarine (E in black and brown) in the Alice's area, and a quantum satellite (in green). Besides, we can see three buoys (red for the submarine A, blue for submarine the B, and black for submarine the E), and a series of rays: the quantum satellite generates pairs of entangled photons which distributes between A and B (orange lines), as well as, receives and transmits classic bits between A and B too (as a part of a classic disambiguation channel, plotted in gray lines). Moreover, we can observe a black line inside the electromagnetic shadow associated to the footprint of the satellite (pink triangular sector) between the not ally submarine E and the quantum satellite. In fact, the Eve's submarine E is close enough to be affected by the electro-magnetic shadow associated to the satellite footprint (pink triangular sector), and far enough not to be detected by Alice, thus being able to decode and thus alter the disambiguation bits of the classic channel. Although the whole figure is out of proportion, it gives us a complete operational notion about this configuration. On the other hand, for this configuration if we resort to a satellite like Micius [25], which orbits at an altitude of ~ 500 km over the Earth, with a classic electromagnetic-type disambiguation channel, the security problem in Fig. 3(a) is shown immediately, given that in these circumstances, this configuration would give us Information security but not Information integrity, e.g., if Alice (A) wants to teleport a qubit to Bob (B) using the protocol in Fig. 2(c) and as part of that protocol it performs a measurement from which two disambiguation bits arise, which it sends to Bob (B) through the classic channel, and this channel is intercepted and decoded by Eve (E), the latter can modify said bits, so Bob (B) will not be able to perform the correct reconstruction of the qubit teleported by Alice (A), although Eve (E) won't be able to either.

It is evident that here there is no alternative, we cannot overcome this serious problem by means of quantum repeaters every 50 km of an optical fiber laying, here we must use a satellite that somehow distributes the entangled photons but not compromise the integrity of Information when an extensive footprint results from satellite antenna's electromagnetic-radiation, which is an essential part of the classic channel. Even less recommended is a geostationary satellite, which is generally located at a height of 30,000 km over the Equator. Since, in that case, the footprint would be even greater and further complicating the problem since it would greatly facilitate the interception capacity of a not ally submarine (E) reducing in equal proportion the possibilities of this eavesdropper to be detected. Unless it was a very special geostationary satellite, which instead of using a classic electromagnetic channel to transmit the disambiguation bits uses another type of link, for example an optical one, with an extremely small solid-angle opening for the transmission of the classic bits and therefore a very small footprint on the surface of the sea. This is the case of the Skyloom-Globe-Corporation's satellite relay called Uhura [8], which will be in orbit in 2022 and whose most relevant technical specifications are: a) 36,000 km over the Equator, b) geostationary, c) over the Atlantic Ocean, d) 2 GigaBits/sec of data transfer, and e) only 1 km footprint on the ocean surface.

This is the case of Fig. 3(b), where the bidirectional optical link (independently of the other link, i.e., the one that distributes the entangled photons) is in yellow. As a direct consequence of this, the action of the not alley submarine (E) is completely out of place thus simultaneously and for the first time preserving the security and integrity of the Information.

On the other hand, in both graphs of Fig. 3, brown connections can be seen between all the buoys and their respective submarines, where these represent a final link of a classic nature, given that in cases A and B, although the submarines are the finally generating and receiving the messages, it is the buoys that carry out the protocols indicated in the red and blue sectors of Fig. 2(c), respectively. Evidently, these connections must be sufficiently resistant to mechanical stretching and compression due to sea agitation on the surface, as well as corrosion due to environmental aspects. This has to do with two aspects: ensuring communication regardless of the weather, and not losing the buoys in the hands of a non-ally.

Finally, unlike an electromagnetic link, clouds as well as certain adverse weather conditions will inevitably interrupt the optical link. These problems are beyond the scope of this work, as well as details of stabilization and visual tracking of the buoys by the satellite, since as we have previously mentioned, the specific focus of this paper has to do with the feasibility study of the protocols on QPU of the IBM Q type [7].

### III. IMPLEMENTATIONS ON A 16-QUBITS QUANTUM PROCESSING UNIT

In this section, we present a couple of metrics need to evaluate the teleportation fidelity for the protocol of Fig. 2(c) and (d), i.e., to evaluate the similarity between the qubit to be teleported and the teleported qubit, which are more appropriated to calculate the quality of teleportation when we work with superconductors platforms like IBM Q [7]. The traditional metric used in optical circuit [6] is known as fidelity [26-28], however, given the characteristics of the outcomes delivered by the IBM Q [7] QPUs in the form of probability bars, it is evident then that new metrics that are designed based on this characteristic will indicate in a much more faithful way the final quality of the teleportation obtained. On the other hand, the implementations of the equivalent bidirectional protocol of Fig. 2(d) on the ibmq_16_melbourne quantum processor [7], with 16 bits, are also presented in this chapter. However, given that the mentioned protocol implies 6 qubits for its implementation, and IBM Q [7] does not present a processor among those of 5 qubits (ibmq_london, ibmq_burlington, ibmq_esre, ibmq_vigo, ibmq_rome, and ibmq_5_yorktown - ibmqx2) and ibmq_16_melbourne, of 16 qubits, we are obliged to use the latter by wasting 10 qubits that are not used.

#### A. Metrics

The following metrics will be defined below, and both have a common denominator, which is that they apply to the probability bars, defined in Eq.(8), of the qubit to be teleported and that which has already been teleported thanks to a QPU or to the IBM Q simulator [7].

*Mean Percentage Absolute Error (MeanPAE):*

The Mean Percentage Absolute Error (MeanPAE) is a quantity which tells us about what percentage linear distance, on average, the outcomes are from the theoretical or ideal values. Therefore, the MeanPAE is given by

$$\text{MeanPAE}(t,o) = \frac{1}{2^{noq}} \sum_{k=0}^{2^{noq}-1} |t_k - o_k| \times 100 \qquad (10)$$

where $t$ is a vector with all the theoric probabilities, $o$ is another vector with all the experimental outcomes (all resulting probabilities), $noq$ is the number of qubits, being $2^{noq}$ the number of probability bars. It is evident that the smaller this value (percentage between o and 100 %), the more faithful the teleportation carried out by the protocol implemented on an IBM Q platform [7] will have been.

*Maximum Percentage Absolute Error (MaxPAE):*

The Maximum Percentage Absolute Error (MaxPAE) represents the worst-case scenario in which the eventual lack of fidelity of a teleportation protocol is exposed in crude when it is implemented on an IBM Q platform [7]. This metric is based on a very simple but important criterion: *every chain has the strength of its weakest link*. This metric is also expressed in percent, being able to vary between 0 and 100 %, and can be represented as follows:

$$\text{MaxPAE}(t,o) = max\{|t_k - o_k| / k = \{0,1,\ldots,2^{noq}-1\}\} \times 100, \tag{11}$$

where $t$ is again a vector containing all the theoric probabilities, and $o$ is another vector containing all the experimental probabilities (outcomes), with $|t_k - o_k| \leq 1$ for $k = \{0,1,\ldots,2^{noq}-1\}$.

If we obtain probabilities from an unique qubit, we will only have $|t_{|0\rangle} - o_{|0\rangle}|$ and $|t_{|1\rangle} - o_{|1\rangle}|$, which result from the following equations:

$$t_{|0\rangle} + t_{|1\rangle} = 1, \text{ and} \tag{12a}$$

$$o_{|0\rangle} + o_{|1\rangle} = 1. \tag{12b}$$

Now, if we subtract, in order, Eq.12(b) from Eq.12(a), it will result,

$$t_{|0\rangle} - o_{|0\rangle} + t_{|1\rangle} - o_{|1\rangle} = 0. \tag{13}$$

Sending $t_{|1\rangle} - o_{|1\rangle}$ to the other side of the equal sign, and applying modulus to both sides of the new equation results,

$$|t_{|0\rangle} - o_{|0\rangle}| = |t_{|1\rangle} - o_{|1\rangle}| = \frac{|t_{|0\rangle} - o_{|0\rangle}| + |t_{|1\rangle} - o_{|1\rangle}|}{2} = max\{|t_{|0\rangle} - o_{|0\rangle}|, |t_{|1\rangle} - o_{|1\rangle}|\} \tag{14}$$

Therefore, under these circumstances, i.e., with probabilities from only one qubit, MeanPAE and MaxPAE will match. Therefore, in the following experiments we will only work with one of the metrics.

Instead, this will not happen with two or more measured qubits. For example, if two qubits are measured simultaneously in IBM Q [7], then, we will have four theoretical probabilities and four ones provided by the QPU:

$$t_{|00\rangle} + t_{|01\rangle} + t_{|10\rangle} + t_{|11\rangle} = 1, \text{ and} \tag{15a}$$

$$o_{|00\rangle} + o_{|01\rangle} + o_{|10\rangle} + o_{|11\rangle} = 1. \tag{15b}$$

Subtracting, in order, Eq.15(a) from Eq.15(b), it results,

$$t_{|00\rangle} - o_{|00\rangle} + t_{|01\rangle} - o_{|01\rangle} + t_{|10\rangle} - o_{|10\rangle} + t_{|11\rangle} - o_{|11\rangle} = 0. \tag{16}$$

Sending $t_{|11\rangle} - o_{|11\rangle}$ to the other side of the equal sign, and applying modulus to both sides of the new equation results,

$$\left|\left(t_{|00\rangle} - o_{|00\rangle}\right) + \left(t_{|01\rangle} - o_{|01\rangle}\right) + \left(t_{|10\rangle} - o_{|10\rangle}\right)\right| = \left|t_{|11\rangle} - o_{|11\rangle}\right|. \tag{16}$$

It is evident that the four individual errors do not have to be equal, in fact, experience indicates that they are not, because each qubit (with which each error is constructed) is generally involved with different gates.

### B. Experiments

The experiments of this subsections are organized in four figures and one table, the first one, Fig. 4 represents both qubits to be employed (THTH, and XTHTH) in the bidirectional quantum teleportation protocol implemented on the circuit composer environment of IBM Q [7], one of its simulation environments together with the simulator itself and qiskit [7]. This figure 4 shows both qubits generated from ground state with its respective metrics, where:

a) it represents the quantum circuit of qubit THTH, very similar to that of Fig. 2(a),

b) the height of the bar is the complex modulus of the wave-function, in this case is 0.9241 for |0>, and 0.3826 = |0.271-0.271j| for |1>, identical to those of Table I (column 2),

c) it is the real part of the state, with 85.3553 % for |0>, and 14.6447 % for |1>,

d) it is the real part density matrix of the state, and

$$\begin{bmatrix} 0.854 & 0.25 \\ 0.25 & 0.146 \end{bmatrix}$$

e) it is the imaginary part density matrix of the state.

$$\begin{bmatrix} 0 & 0.25 \\ -0.25 & 0 \end{bmatrix}$$

f) it represents the quantum circuit of qubit XTHTH, very similar to that of Fig. 2(b),

g) the height of the bar is the complex modulus of the wave-function, in this case is 0.383 for |0>, and 0.9234 = |0.653-0.653j| for |1>, identical to those of Table I (column 3),

h) it is the real part of the state, with 14.6447 % for |0>, and 85.3553 % for |1>,

i) it is the real part density matrix of the state, and

$$\begin{bmatrix} 0.146 & 0.25 \\ 0.25 & 0.854 \end{bmatrix}$$

j) it is the imaginary part density matrix of the state.

$$\begin{bmatrix} 0 & -0.25 \\ 0.25 & 0 \end{bmatrix}$$

Figure 5 represents the implementation of both qubits of Fig. 4 both on the simulator and on the ibmq_armonk processor (QPU) of IBM Q [7] of 1 qubit, where:

a) represents the connectivity map and error rate for two gates of the QPU,

*qubit THTH:*

b) are the Probabilities (outcomes) on the simulator,

c) are the Probabilities (outcomes) on the QPU,

d) represents the Run details of the QPU (fairshare Run mode, and 1024 shots),

*qubit XTHTH:*

e) are the Probabilities (outcomes) on the simulator,

f) are the Probabilities (outcomes) on the QPU,

g) represents the Run details of the QPU (fairshare Run mode, and 1024 shots).

We can see a marked difference between the outcomes of both qubits on the QPU with respect to their theoretical values in Fig. 4, even with those simulator values in Fig. 5. This is due to decoherence [13], however, in this case, there are few gates involved, so we must monitor the outcomes when we implement the protocol on the QPU, trying to compensate (as far as IBM allows) for unfavorable results as a result of decoherence, increasing the number of shots. In fact, if we use the metrics defined above, the percentage errors for the qubits in Fig. 5 were as follows:

for the qubit THTH:

MeanPAE = MaxPAE =1.5663 %,

for the qubit XTHTH:

MeanPAE = MaxPAE = 5.8633 %,

That is to say, if we want this type of platform to be representative to be able to test the viability of the bidirectional quantum teleportation protocol of Fig. 2(d), which involves a lot of quantum gates, we must significantly lower the mentioned percentage errors.

Next we implement the protocol of Fig. 2(d) on the different environments available on the IBM Q [7] platform (circuit composer, simulator, and the ibmq_16_melbourne quantum processor or QPU), but separating the experiment from the point of view of quantum measurements carried out at both ends of the protocol in order to obtain clearer outcomes. Then, Fig. 6 shows:

a) the quantum circuit on IBM Q [7], where, the qubit THTH to be teleported from the left side of q[0] to the right side to q[5] is measured on the latter, and based on this measurement we will obtain the following outcomes,

b) the height of the bar is the complex modulus of the wave-function in the circuit composer,

c) the real part of the state in the circuit composer,

d) Probabilities (outcomes) on the simulator,

e) Probabilities (outcomes) on the QPU,

f) connectivity map of the ibmq_16_melbourne quantum processor (QPU), and

g) the Run details of the QPU (fairshare Run mode, and 8192 shots).

While, Fig. 7 shows:

a) the quantum circuit on IBM Q [7], where, the qubit XTHTH to be teleported from the left side of q[5] to the right side to q[0] is measured on the latter, and based on this measurement we will obtain the following outcomes,

b) the height of the bar is the complex modulus of the wave-function in the circuit composer,

c) the real part of the state in the circuit composer,

d) Probabilities (outcomes) on the simulator,

e) Probabilities (outcomes) on the QPU, and

f) the Run details of the QPU (fairshare Run mode, and 8192 shots).

Table II shows that in both figures, we have significantly decreased both percentage errors for both qubits (in fact, by half) by considerably increasing the number of shots, from 1024 in the case of Fig. 5, to 8192 of the Figures 6 and 7. This gives us a margin of reliability considering that by increasing the number of shots even further we could continue to improve the metrics and thus make the study of this protocol more representative on a platform such as that of IBM Q [7].

## IV. CONCLUSIONS AND FUTURE WORKS

### A. Conclusions

The experiments of Sec. III performed on the ibmq_16_melbourne quantum processor of IBM Q [7] thanks to a modified version of the original bidirectional quantum teleportation protocol of Fig. 2(c), allow us to evaluate the viability, and robustness (i.e., immunity to noise) of the mentioned protocol at a cost practically equal to zero although we had to increase the number of shots eight times to be able to halve the percentage errors. We could not further improve outcomes due to 8192 is the maximum number of shots allowed by IBM Q [7] for this family of QPUs. In fact, it is all that can be done with this type of processors based on superconductor technology [7], confined in an adiabatic environment in order not to lose the coherence of the elements involved, i.e., test the protocol. Specifically, decoherence [13] by loss of adiabaticity would automatically trigger the destruction of the entanglement [22-24] in all the elements that were using it, as well as the loss of coherence of each qubit. Therefore, a real implementation of the aforementioned protocol on a satellite such as the Uhura of Skyloom Globe Corporation [8] is exclusively optical. With the additional possibility offered by this satellite platform, which is that the classic disambiguation channel is also optical, making all the configuration provide security and integrity of the keys exchanged between both allied submarines.

Finally, the solution presented in this work is absolutely reproducible between aircrafts, ground units, or even more interesting, between the Moon and the Earth.

### B. Future Works

Future lines of research in this area will necessarily involve protocols for bidirectional quantum teleportation based literally on the teleportation of entangled photons as a more efficient alternative to entanglement swapping [29-35], especially in a spatial context, with the idea of taking quantum

Internet [36-41] to the Moon and later to Mars, something that is quite improbable with current satellite quantum repeaters configurations [42-44].

**Data Availability**

The experimental data that support the findings of this study are available in ResearchGate with the identifier https://doi.org/10.13140/RG.2.2.21596.62087.

**Acknowledgements**

M. Mastriani thanks to Marcos Franceschini, CEO of Skyloom Global Corporation for his support and permanent predisposition to answer all our queries. We gratefully acknowledge the IBM-Q team for providing us with access to their 16-qubit platform. The views expressed are those of the authors and do not reflect the official policy or position of IBM or the IBM Quantum Experience team.

**Competing interests**

Authors declare they has no competing interests.

**Author Contributions**

SSI is responsible for the project's conceptualization, and management, which concludes in this paper. The effort was planned and supervised by SSI and co-supervised by KJKL. MM designed the study, conceived the protocols and the satellite configurations, designed the quantum circuits, performed the experiments, and wrote the first version of the paper. SSI and KJKL analyzed the results. SSI and KJKL reviewed the first version of the paper. SSI and KJKL wrote the final version of the paper. All authors read and approved the final manuscript.

**Funding**

This research received no external funding.

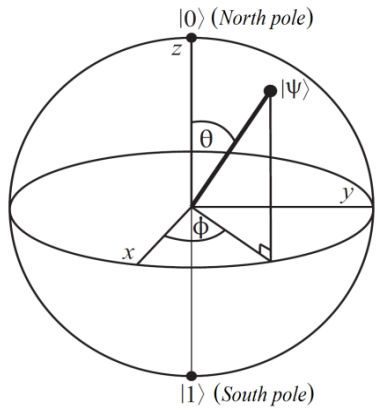

FIG. 1: Bloch's Sphere.

Table I. Metrics' values for qubits THTH and XTHTH.

| Metrics | THTH | XTHTH |
|---|---|---|
| *statevector* | [ 0.924+0j, 0.271-0.271j ] | [ 0.383+0j, 0.653+0.653j ] |
| *amplitude* | 0.924 | 0.383 |
| *phase angle* | 0 | 0 |
| *real part of density matrix* | $\begin{bmatrix} 0.85 & 0.25 \\ 0.25 & 0.15 \end{bmatrix}$ | $\begin{bmatrix} 0.15 & 0.25 \\ 0.25 & 0.85 \end{bmatrix}$ |
| *imaginary part of density matrix* | $\begin{bmatrix} 0 & 0.25 \\ -0.25 & 0 \end{bmatrix}$ | $\begin{bmatrix} 0 & -0.25 \\ 0.25 & 0 \end{bmatrix}$ |
| Po\|0> | 85.35534 % | 14.64466 % |
| Po\|1> | 14.64466 % | 85.35534 % |

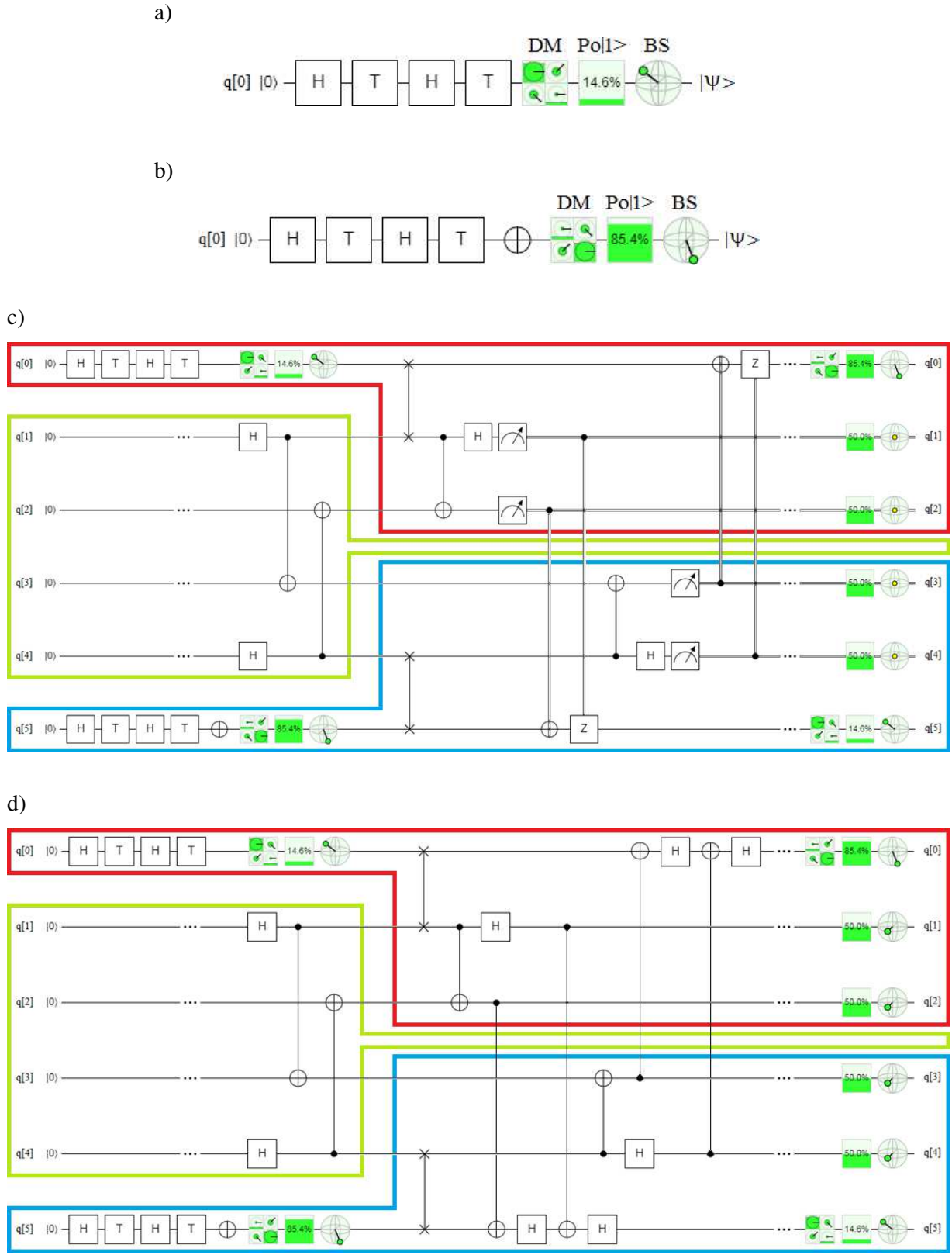

FIG. 2: a) qubit THTH generated from ground state |0> with its respective metrics on Quirk [17]: density matrix (DM), Probability of |1> (Po|1>), and Bloch's sphere (BS); b) idem to previous case but for qubit XTHTH; c) bidirectional teleportation protocol, with *if-then-else* statement, on Quirk [17] between Alice (red on sea) and Bob (blue on sea) thanks to a quantum satellite (green); and d) alternative bidirectional teleportation protocol, without *if-then-else* statement, on Quirk [17] between Alice (red on sea) and Bob (blue on sea) thanks to a quantum satellite (green).

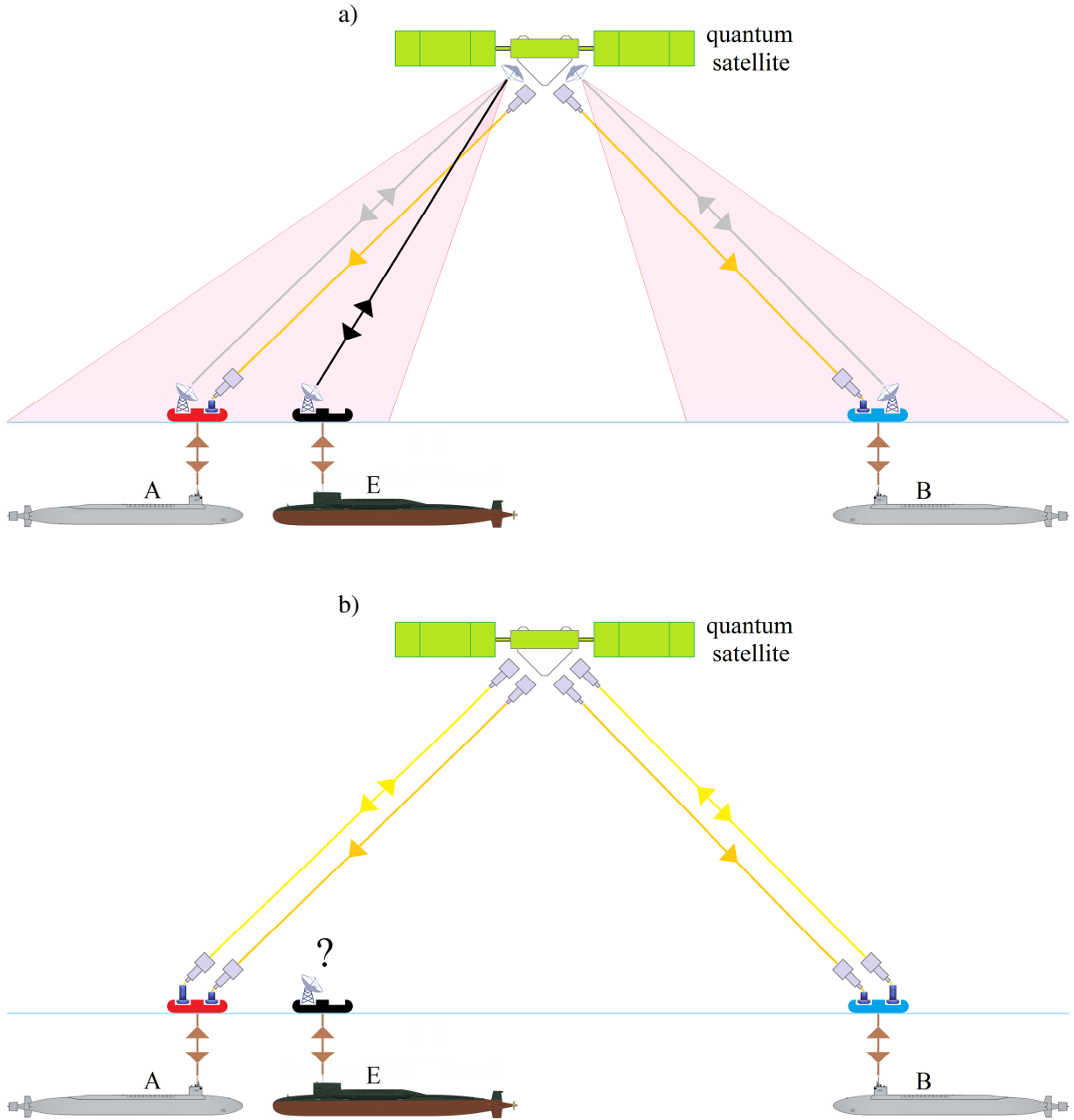

FIG. 3: A satellite (green) to communicate Alice (a red buoy on sea in point A) and Bob (a blue buoy on sea in point B), where the color code respect to Figs. 1(c) and (d) was respected. Besides, there is a third ship, the Eve's submarine, silent, submerged and pre-existing in the Alice's area. Therefore, two alternatives emerge: a) the Eve's submarine is at an appropriate distance from Alice's one, i.e., close enough to be affected by the electromagnetic shadow associated to the satellite footprint (pink triangular sector), and far enough not to be detected by Alice, thus being able to decode and thus alter the disambiguation bits of the classic channel, instead, b) using a satellite like Skyloom's [8] Uhura with a fully optical classic channel, which can focus exclusively on Alice's buoy for transmission and reception of classic disambiguation bits, Eve's submarine has no chance of altering the communication between Alice and Bob. Otherwise, the orange rays in (a) and (b) represent the entangled photons scattered across the satellite, while the brown rays represent the cables between the buoys and the submarines, which are subjected to great forces of stretching and compression, as well as mechanical degradation due to exposure to the environment. Moreover, in (a), the gray rays represent the electromagnetic links that drives the transmission of the classic bits of disambiguation that protocol needs to rebuild the teleported states, while the black ray represents the Eve's intervention in the electromagnetic channel, whereas, in (b) the yellow ray represents the optical link thanks to Uhura. They are the Alice's and Bob's buoys that reconstruct the teleported states and emit them to the submarines. Finally, all the elements of the figures are out of proportion in order to make them more visible.

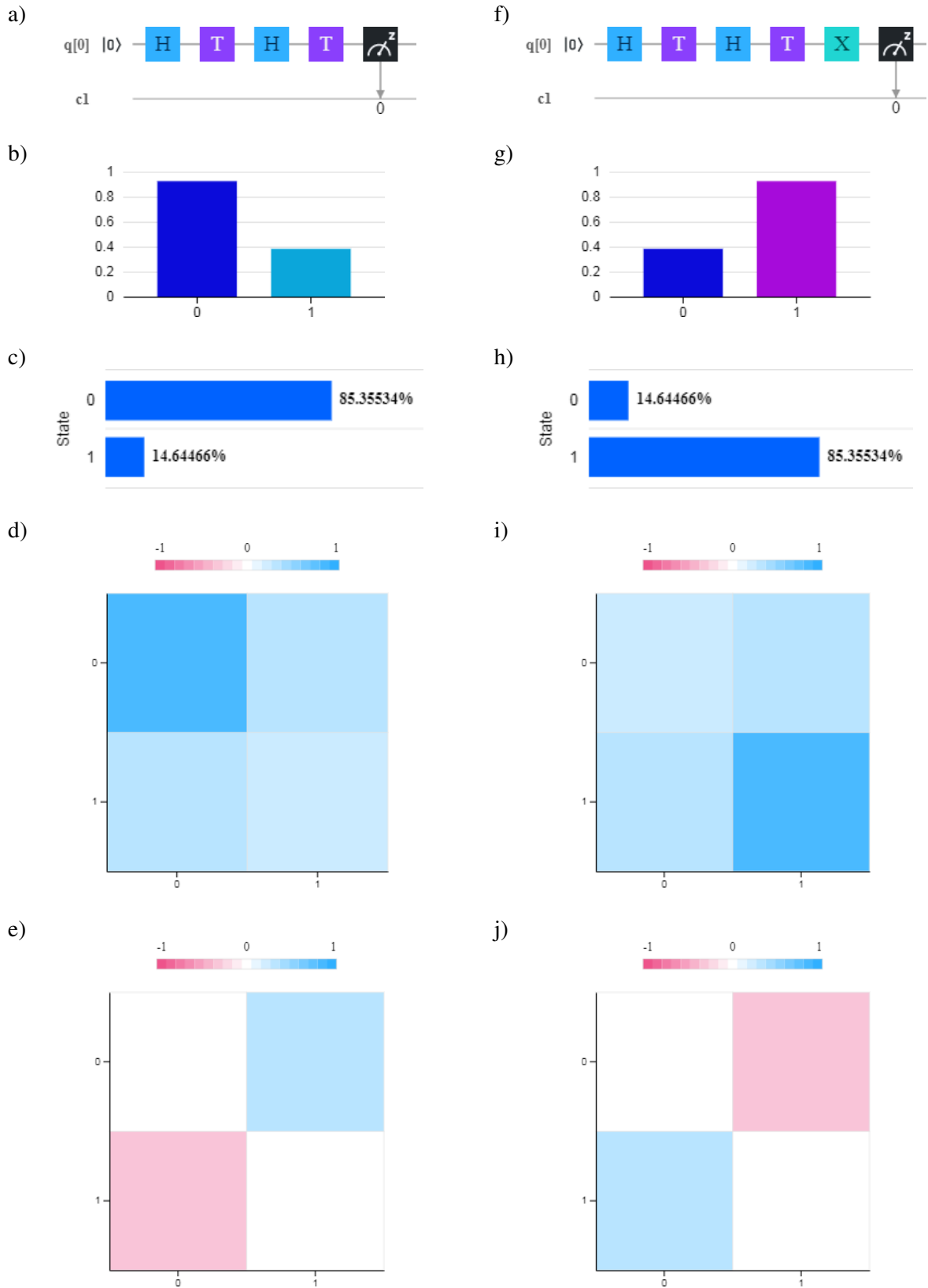

FIG. 4: Qubit THTH generated from ground state with its respective metrics on the circuit composer of IBM Q [7]: a) quantum circuit, b) height of the bar is the complex modulus of the wave-function, c) real part of the state, (d) real part density matrix of the state, and e) imaginary part density matrix of the state. Qubit XTHTH generated from ground state with its respective metrics on the circuit composer of IBM Q [7]: f) quantum circuit, g) height of the bar is the complex modulus of the wave-function, h) real part of the state, i) real part density matrix of the state, and j) imaginary part density matrix of the state.

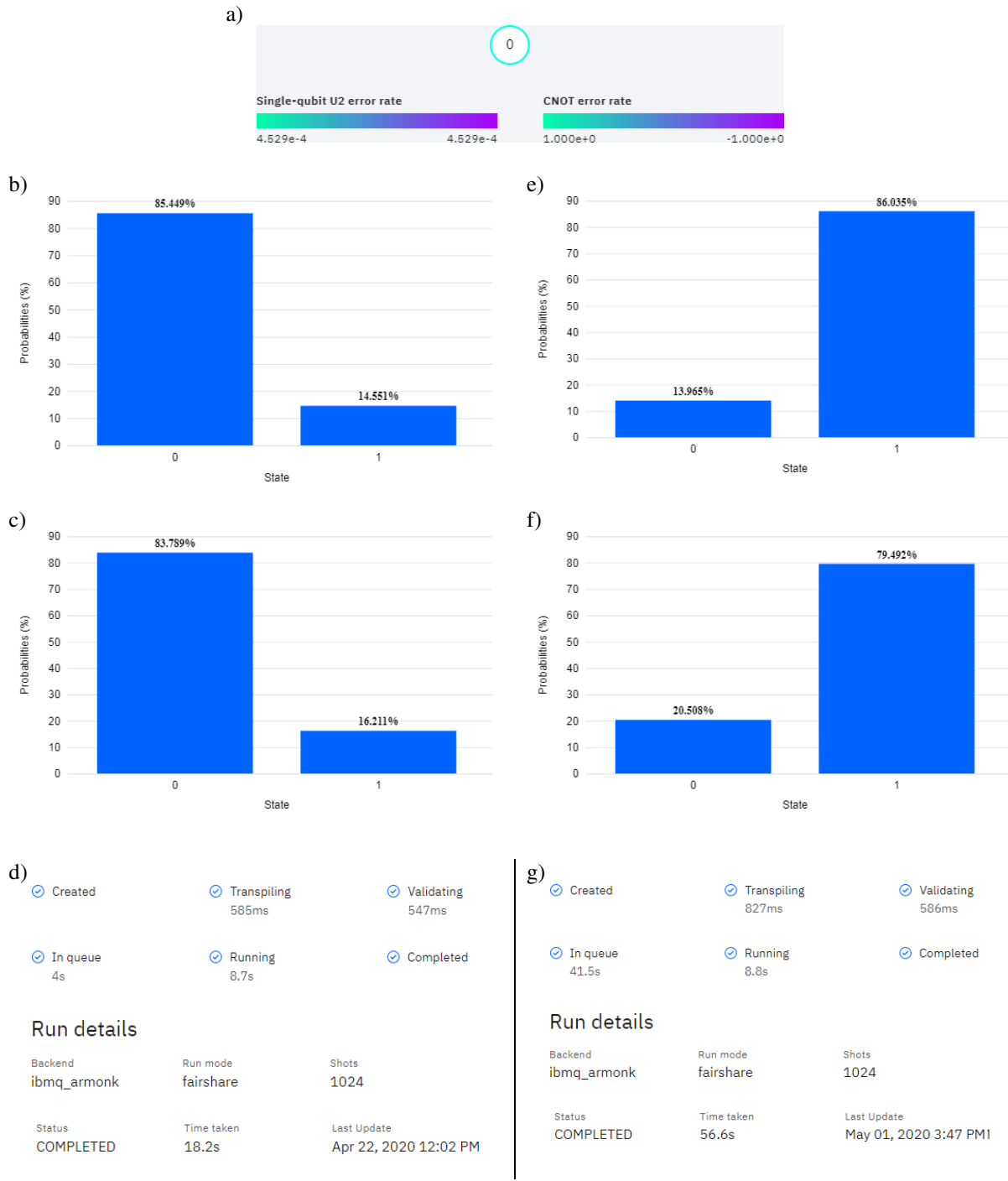

FIG. 5: a) connectivity map and error rate for two gates of ibmq_armonk quantum processor (QPU) of IBM Q [7], while the rest represents the metrics of both qubits, for the simulator and the QPU. Qubit THTH: b) Probabilities (outcomes) on the simulator, c) Probabilities (outcomes) on the QPU, and d) Run details of the QPU. Qubit XTHTH: e) Probabilities (outcomes) on the simulator, f) Probabilities (outcomes) on the QPU, and g) Run details of the QPU.

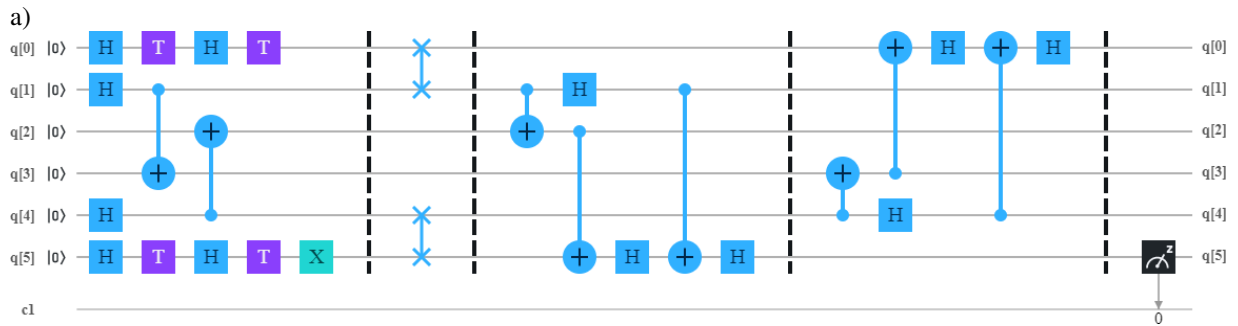

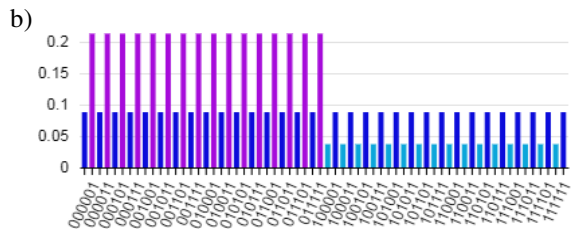
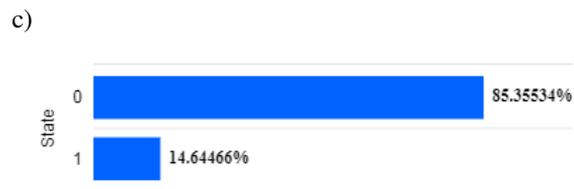

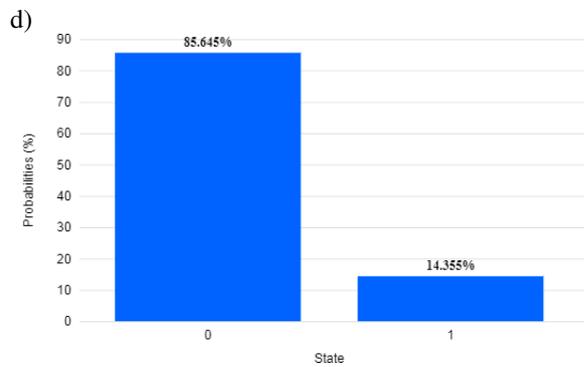
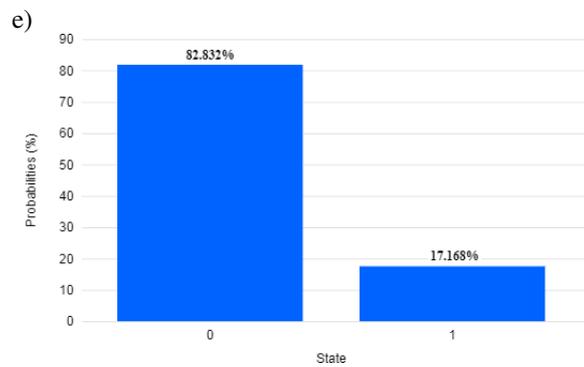

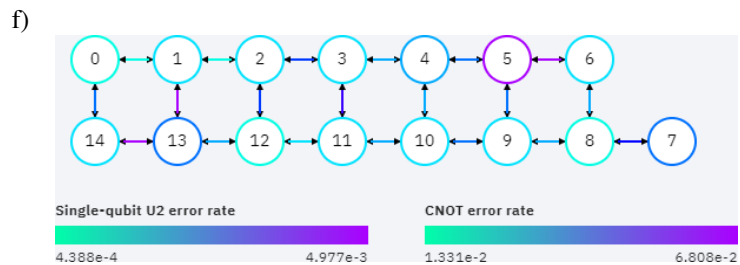

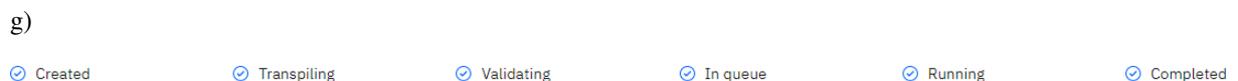

FIG. 6: a) bidirectional teleportation protocol (quantum circuit) on IBM Q [7] with quantum measurement in qubit q[5], b) height of the bar is the complex modulus of the wave-function in the circuit composer, c) real part of the state in the circuit composer, d) Probabilities (outcomes) on the simulator, e) Probabilities (outcomes) on the QPU, f) connectivity map of the ibmq_16_melbourne quantum processor (QPU), and g) Run details of the QPU.

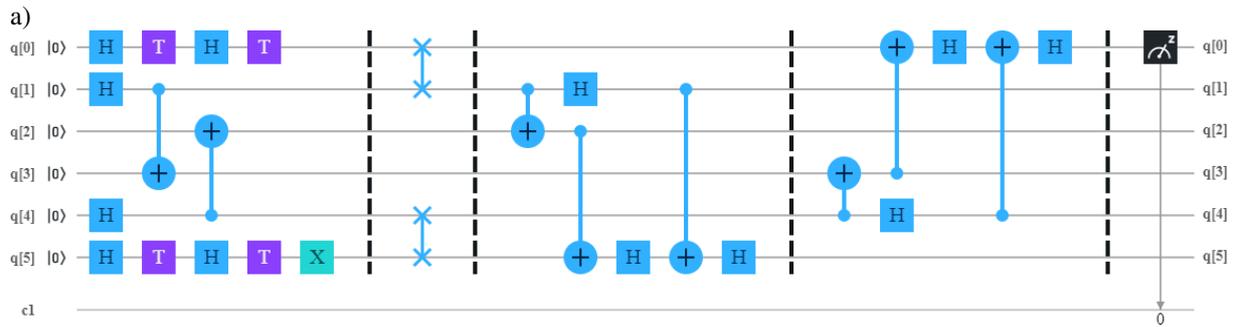

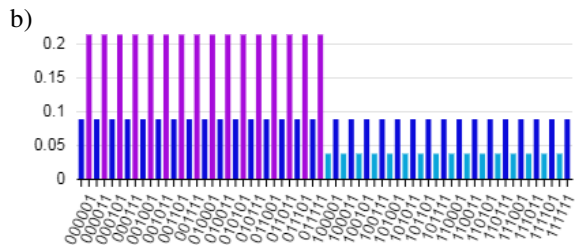
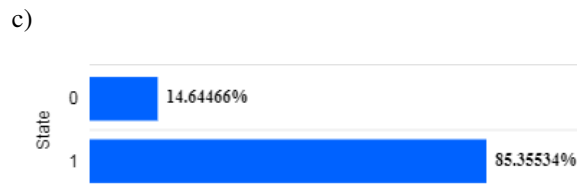

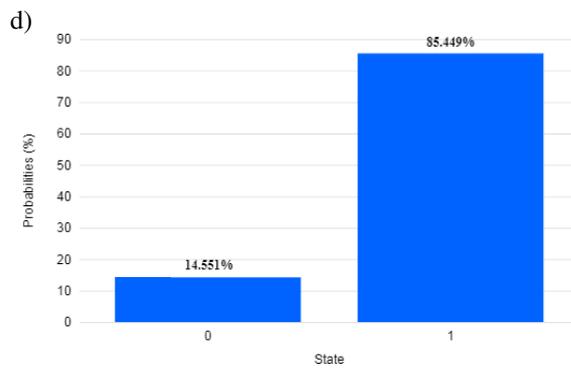
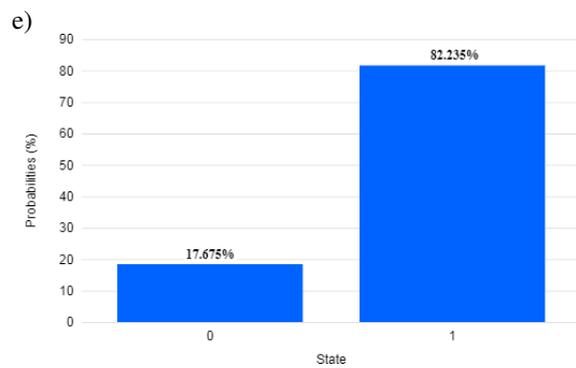

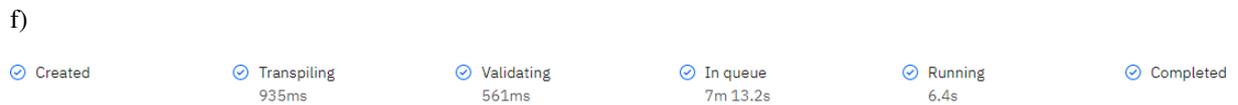

FIG. 7: a) bidirectional teleportation protocol (quantum circuit) on IBM Q [7] with quantum measurement in qubit q[0], b) height of the bar is the complex modulus of the wave-function in the circuit composer, c) real part of the state in the circuit composer, d) Probabilities (outcomes) on simulator, e) Probabilities (outcomes) on the ibmq_16_melbourne quantum processor (QPU), and f) Run details of the QPU.

Table II. Metrics' values for qubits THTH and XTHTH.

| Qubit | Figures | Metrics (MeanPAE = MaxPAE) |
|---|---|---|
| THTH | 6 (c) Theoric, and 6 (d) Simulator | 0.2897 |
| THTH | 6 (c) Theoric, and 6 (e) QPU | 2.5233 |
| XTHTH | 7 (c) Theoric, and 7 (d) Simulator | 0.0937 |
| XTHTH | 7 (c) Theoric, and 7 (e) QPU | 3.0753 |